\documentclass[prl,twocolumn,preprintnumbers,amsmath,amssymb]{revtex4}

\usepackage{latexsym}
\usepackage{hyperref}

\newcommand{\SL}{\mbox{SL}(2,\mathbb{R})}
\newcommand{\re}[1]{(\ref{#1})}

\def\eps{\epsilon}

\def\d{\partial}

\def\cK{{\cal K}}
\def\cL{{\cal L}}

\def\cQ{{\cal Q}}

\def \eps{{\epsilon}}

\begin{document}
\preprint{IFUM-928}
\title{Semi-classical central charge in topologically massive gravity}

\author{Geoffrey Comp\`ere}
\affiliation{%
Department of Physics, University of California at Santa Barbara, \\ Santa Barbara, CA 93106, USA
}

\author{St\'ephane Detournay}

\affiliation{%
Istituto Nazionale di Fisica Nucleare,\\
Via Celoria 16, 20133 Milano, Italy}

\begin{abstract}
It is shown that the warped black holes geometries discussed recently in 0807.3040 admit an algebra of asymptotic symmetries isomorphic to the semi-direct product of a Virasoro algebra and an algebra of currents. The realization of this asymptotic symmetry by canonical charges allows one to find the central charge of the Virasoro algebra.  The right-moving central charge $c_R = \frac{(5\hat{\nu}^2+3)l}{G\hat{\nu} (\hat{\nu}^2+3)}$ is obtained when the Virasoro generators are normalized in order to have a positive zero mode spectrum for the warped black holes. The current algebra is also shown to be centrally-extended. 
\end{abstract}


\maketitle

There has been recently a lot of activity around the so-called topologically massive gravity (TMG) theory \cite{Deser:1981wh,Deser:1982vy} whose action reads
\begin{equation}
 I_{TMG} = \frac{1}{16 \pi G} \left[\int_M  d^3x \, \sqrt{-g} (R+ \frac 2 {l^2}) + \frac{1}{\mu} \; I_{CS} \right].
\end{equation}
The gravitational Chern-Simons term $I_{CS}$ is given by
\begin{equation} 
 I_{CS} = \frac{1}{32 \pi G} \int_M d^3x \, \sqrt{-g} \varepsilon^{\lambda \mu \nu} \Gamma^\alpha_{\lambda \sigma} \left(\partial_\mu \Gamma^\sigma_{\alpha \nu} + \frac{2}{3} \Gamma^{\sigma}_{\mu \tau}\Gamma^{\tau}_{\nu \alpha}\right),
 \end{equation}
where $G$ is the Newton's constant, $l$ is the $adS_3$ curvature and $\mu$ is the Chern-Simons coupling which can be taken positive without loss of generality. 

Any solution of Einstein's gravity with negative cosmological constant is automatically a solution of TMG. Therefore, $adS_3$ constitutes a candidate for the vacuum of the theory. The analysis of the asymptotic symmetry algebra for asymptotically $adS_3$ space-times in the Brown-Henneaux approach reveals the appearance of two Virasoro algebras with different central charges in the left and right sectors \cite{Hotta:2008yq} (see also \cite{Kraus:2005zm,Solodukhin:2005ah,Park:2006gt} ):
\begin{equation}\label{cAdSTMG}
 c_{L/R} = \frac{3 l}{2 G} \left( 1 \pm \frac{1}{\mu \,l} \right).
\end{equation}
For the BTZ black holes \cite{Banados:1992gq}, whose mass and angular momentum in TMG are given by $M_{BTZ}= M - J/(\mu l^2)$ and $J_{BTZ} = J - M /\mu$ in terms of their mass $M$ and angular momentum $J$ in Einstein gravity \cite{Garcia:2003nm,Moussa:2003fc}, \re{cAdSTMG} leads to an agreement between the macroscopic entropy computed from the  Iyer-Wald-Tachikawa formula \cite{Iyer:1994ys,Tachikawa:2006sz} and that obtained by applying the Cardy formula \cite{Solodukhin:2005ah,Hotta:2008yq}. This points to an interesting dual CFT description. However, TMG seems to suffer from some pathologies, and it is not clear yet how these could be dealt with. For $G>0$, indeed, the massive excitations about $adS_3$ carry negative energy \cite{Deser:1982vy,Deser:1981wh}. In that case, $adS_3$ is not a stable vacuum. Flipping the sign of $G$ would allow to cure the problem, but will in turn yield a negative mass for the BTZ black holes. This makes sense if there is a superselection rule forbidding the BTZ solutions \cite{Carlip:2008jk,Carlip:2008eq}. The special case $\mu \, l = 1$ raises numerous questions well beyond the scope of this paper \cite{Li:2008dq,Strominger:2008dp,Carlip:2008qh,Grumiller:2008qz,Carlip:2008eq,Carlip:2008jk,Giribet:2008bw,Park:2008yy}.

In this note, we shall focus on generic values for the Chern-Simons coupling $\mu$ and $G >0$. In that case, it has been suggested that the theory could display other stable backgrounds around which it could be expanded, named spacelike, timelike or lightlike \emph{warped anti-de Sitter spaces} \cite{Anninos:2008fx}. Spacelike warped adS spacetimes admit the left-broken isometry group $U(1)_L \otimes SL(2,\mathbb R)_R$. Interestingly, there exist black hole solutions which can be obtained by performing discrete identifications in this background (such solutions have appeared in various other contexts, see references in \cite{Anninos:2008fx}). 

For $\mu \, l > 3$, the warped adS background is said to be \emph{stretched} and the black holes ( discussed already in \cite{Nutku:1993eb,Gurses:1994aa}, see also \cite{Moussa:2008sj}) are regular. 
For $\mu \, l < 3$, instead, the background is said to be \emph{squashed} and the quotient yields closed time-like curves. The corresponding black holes can be identified with the so-called three-dimensional G\"odel black holes \cite{Banados:2005da}. The latter geometries also represent solutions to Einstein-Maxwell-Chern-Simons theory \cite{Banados:2005da} and are part of an exact string theory background through a marginal deformation of the $\SL$ WZW model supplemented by the appropriate background fields \cite{Israel:2004vv,Detournay:2005fz,Compere:2008cw}. On the other hand, the spacelike stretched black holes are formal solutions of the same theories, but have to be supported by imaginary matter fields.

Based on various heuristic arguments, it has been conjectured in \cite{Anninos:2008fx} that TMG defined with suitable ``warped adS'' boundary conditions would be dual to a two-dimensional CFT with central charges given by 
\begin{eqnarray}
 c_R &=& \frac{(5 {\hat{\nu}}^2 + 3)l}{G {\hat{\nu}}({\hat{\nu}}^2 + 3)}, \qquad  c_L = \frac{4 {\hat{\nu}} l}{G ({\hat{\nu}}^2 + 3)}
\end{eqnarray}
with ${\hat{\nu}} = \mu \,l/3$. We would like to investigate whether these central extensions could be derived from the asymptotic symmetry algebra associated with spacelike warped $adS$ geometries.

\section{Asymptotically conserved charges in Topologically Massive Gravity}

The conserved charges for linearized TMG around a fixed anti-de Sitter background were found in \cite{Deser:2003vh,Deser:2005jf}. In order to compute charges associated with Killing vectors around warped adS geometries, the more general expression found in \cite{Bouchareb:2007yx} is needed. In that reference, the conserved charges are computed in the linearized TMG around an arbitrary background $g_{\mu\nu}$ for Killing vectors $\xi$. Asymptotic Killing vectors are defined as diffeomorphisms which are not gauge transformations, i.e. which generate non-trivial charges. For asymptotic Killing vectors and if the Killing equations do not fall-off fast enough close to the boundary, additional contributions may appear in the asymptotic charges. A more general analysis is therefore needed. 

The conserved charges associated with an asymptotic Killing vector $\xi$ can be derived following the procedure outlined in \cite{Barnich:2003xg,Barnich:2007bf}. In that framework, the charge differences $\cQ_\xi[g,\bar g]$ between the reference solution $\bar g$ and the solution of interest $g$ are defined as
\begin{equation}
\cQ_\xi[g ;\bar g] = \int_{\bar g}^g \int_S \sqrt{-g}\, k^{\mu\nu}_\xi[\delta g ; g] \eps_{\mu\nu\rho}\,dx^\rho
\end{equation}
where the first integration is performed in the phase space of solutions, $S$ is the sphere at infinity and $k^{\mu\nu}[\delta g,g]\eps_{\mu\nu\rho}dx^\rho$ is a one-form defined in the linearized theory. The charges do not depend on the path chosen in the integration if the integrability condition $ \delta\,\oint_S k^{\mu\nu}_\xi[\delta g ; g] \eps_{\mu\nu\rho}\,dx^\rho = 0$ holds. The crucial property of these charges is that they represent the Lie algebra of asymptotic symmetries via a covariant Poisson bracket up to central charges
\begin{equation}
\{\cQ_\xi[g ;\bar g],\cQ_{\xi^\prime}[g ;\bar g]\} = \cQ_{[\xi,\xi^\prime]}[g ;\bar g] + \cK_{\xi,\xi^\prime}[\bar g].
\label{eq:repr}
\end{equation}
Here, $[\xi,\xi^\prime]$ is the Lie bracket and $\cK_{\xi,\xi^\prime}[\bar g] \equiv \int_S k^{\mu\nu}_{\xi}[\cL_{\xi^\prime} \bar g;\bar g]$. More precisely this result holds modulo a technical assumption (see (4.3) of \cite{Barnich:2007bf}) which will be checked for the case at hand. 

A similar representation theorem holds \cite{Koga:2001vq} for the charges defined with covariant phase space methods \cite{Iyer:1994ys}. However, this theorem has not been generalized to our knowledge to non-diffeomorphic invariant actions like TMG. The canonical Hamiltonian approach \cite{Brown:1986ed,Brown:1986nw} developed for TMG in \cite{Hotta:2008yq} is equivalent to our approach as proven e.g. in \cite{Barnich:2007bf}.

The only quantity to be computed is the conserved form $k^{\mu\nu}_{\xi}[\delta g;g]$ of the linearized TMG theory. From the uniqueness results \cite{Barnich:2001jy}, it is clear that this charge should reduce on-shell to the one found in \cite{Bouchareb:2007yx} when $\xi$ is a Killing vector. Denoting $\delta g_{\mu\nu} \equiv h_{\mu\nu}$, we can therefore write \footnote{ We differ by a conventional overall minus sign with respect to \cite{Bouchareb:2007yx}.}
\begin{eqnarray}
(16 \pi G)\,k^{\mu\nu}_{\xi}[\delta g;g] &=& (16 \pi G)\,k^{\mu\nu}_{Ein,\xi_{tot}}[\delta g;g] \nonumber\\
&& \hspace{-2cm} -\frac{1}{\mu\sqrt{-g}}\xi_\lambda \left( 2 \eps^{\mu\nu\rho}\delta (G^\lambda_{\;\rho}) -  \eps^{\mu\nu\lambda}\delta G )\right) \nonumber \\
 && \hspace{-2cm}-\frac{1}{\mu\sqrt{-g}} \eps^{\mu\nu\rho} \left( \xi_\rho h^{\lambda\sigma}G_{\sigma\lambda} +\frac 1 2 h(\xi_\sigma G^\sigma_{\;\rho}+\frac 1 2 \xi_\rho R) \right)\nonumber \\
&& +(16 \pi G)\,E^{\mu\nu}[\delta g ;\cL_\xi g],\label{chargetot}
\end{eqnarray}
where $\xi_{tot}^{\nu} = \xi^\nu + \frac 1 {2\mu \sqrt{-g}} \eps^{\nu \rho\sigma}D_\rho \xi_\sigma$, $k^{\mu\nu}_{Ein,\xi}[\delta g;g]$ is the Iyer-Wald expression \cite{Iyer:1994ys} for general relativity
\begin{eqnarray}
(16 \pi G)\sqrt{-g}\,k^{\mu\nu}_{Ein,\xi}[\delta g;g] &=& \sqrt{-g}\,\xi^\mu(D_\lambda h^{\lambda \nu}-D^\lambda h) \nonumber \\
&&\hspace{-20pt} -\delta(\sqrt{-g}D^\mu \xi^\nu) - (\mu \leftrightarrow \nu)
\end{eqnarray}
where the variation acts only on $g$ and $E^{\mu\nu}[\delta g ;\cL_\xi g]$ is an additional contribution linear in the Killing equation and its derivatives.

The term $E^{\mu\nu}[\delta g ;\cL_\xi g]$ can be obtained as follows. One defines the on-shell vanishing Noether current $S_\xi^\mu[g] = 2 \frac{\delta L}{\delta g_{\mu\nu} }\xi_\nu = - \frac{\sqrt{-g}}{8\pi G}(G^{\mu\nu}+\Lambda g^{\mu\nu}+\frac{1}{\mu}C^{\mu\nu})\xi_\nu$ associated with diffeomorphisms. Using several integrations by parts, the variation of the Noether current can be decomposed as follows
\begin{eqnarray}
\delta S_\xi^\mu[g] &=& \delta g_{\alpha \beta} \frac{\delta L}{\delta g_{\alpha\beta}}\xi^\mu \nonumber \\
&&+ W^\mu[\delta g ; \cL_\xi g] + \sqrt{-g}D_\nu k_\xi^{\mu\nu}[\delta g ;g]. \label{decomp}
\end{eqnarray}
The left-hand side and the first term on the right-hand side vanish for solutions $g$ and linearized solutions $\delta g$ while the symplectic structure $W^\mu[\delta g ; \cL_\xi g]$ vanishes for Killing vectors. The superpotential $ k_\xi^{\mu\nu}$ is therefore conserved under these conditions. The decomposition \eqref{decomp} is however not unique, even on-shell, as one can freely add two compensating terms proportional to the Killing equations to $W^\mu$ and $\sqrt{-g}D_\nu  k_\xi^{\mu\nu}$.

The conserved charge $k_\xi^{\mu\nu}[\delta g ;g]$ of \cite{Barnich:2001jy,Barnich:2007bf} is defined via a contracting homotopy acting on the Noether current which provide a definite prescription to perform the decomposition \eqref{decomp}. Roughly, one performs the integrations by parts without using the Killing equations to simplify the total derivative $\sqrt{-g} D_\nu  k_\xi^{\mu\nu}[\delta g ;g]$. It is then an exercise to redo the explicit computation of \cite{Bouchareb:2007yx} keeping all terms in $k_\xi^{\mu\nu}[\delta g ;g]$. We obtain \eqref{chargetot} with
\begin{eqnarray}
(16 \pi G)E^{\mu\nu}[\delta g ;\cL_\xi g] &=& \frac 1 2 \left( h^\nu_{\;\,\lambda}\cL_\xi g^{\lambda \mu} - h^\mu_{\;\,\lambda}\cL_\xi g^{\lambda \nu}\right) \nonumber\\&& 
\hspace{-2cm}+\frac{1}{4\mu \sqrt{-g}}\eps^{\mu\nu\rho}\left(
D_\lambda \cL_\xi g^\lambda_{\;\;\rho}- D_\rho \cL_\xi g_\lambda^{\;\;\lambda}  
  \right) h.\label{defE}
\end{eqnarray}
The first term can be recognized as the supplementary contribution to the Iyer-Wald expression for the conserved charges in general relativity that should be added to recover the expression derived in \cite{Abbott:1981ff,Anderson:1996sc,Barnich:2001jy}.

\section{Asymptotic symmetry algebra of spacelike warped anti-de Sitter}

Let us start with the black hole geometries in the spacelike deformation of anti-de Sitter described in \cite{Anninos:2008fx},
\begin{eqnarray}\label{BH}
\frac{ds^2}{l^2} &=& dt^2 + \frac{dr^2}{(\hat{\nu}^2+3)(r - r_+)(r-r_-)} \\
&& +(2\hat{\nu} r-\sqrt{r_+ r_-(\hat{\nu}^2+3)})dt \;d \theta +\frac{r}{4}\{ 3(\hat{\nu}^2-1) r \nonumber\\
&&  +(\hat{\nu}^2+3)(r_+ +r_-)-4\hat{\nu} \sqrt{r_+ r_-(\hat{\nu}^2+3)}\}d\theta^2 .\nonumber
\end{eqnarray}
We first observe that the squashed metrics $\hat{\nu}^2 < 1$ are exactly the G\"odel black holes described in \cite{Banados:2005da} 
\begin{eqnarray}\label{BH2}
ds^2 &=& dT^2  +(-\frac{4 G J}{\alpha}+8G \nu R-\frac{2}{L^2}(1-\alpha^2 L^2)R^2)d\Phi^2 \nonumber \\
&&- 4 \alpha R dT d\Phi +\frac{dR^2}{\frac{4 G J}{\alpha} -8G \nu R +\frac{2(1+\alpha^2 L^2)}{L^2}R^2}
\end{eqnarray}
up to a change of coordinates. The relation between the anti-de Sitter radius $L$ and the Chern-Simons coupling $\alpha$ of \cite{Banados:2005da} and the TMG parameters $(\hat{\nu},l)$ is given by
\begin{eqnarray}
\hat{\nu}^2& =& \frac{3\alpha^2 L^2}{2+\alpha^2 L^2},\qquad  \qquad  l^2 = \frac{3 L^2}{2+\alpha^2 L^2}.
\end{eqnarray}
A detailed study of the relation between $3d$ G\"odel geometries and warped black holes is given in the companion paper \cite{Compere:2008cw}. The range of parameters in which there are closed timelike lines $\hat{\nu}^2 < 1$ corresponds to the range $\alpha^2 l^2 < 1$ where the metric is solution of Einstein-Maxwell-Chern Simons theory. In TMG, the solutions \eqref{BH} exist also for the streched case $\hat{\nu}^2 \geq 1$ and are well-behaved because of the lack of closed timelike lines. We now restrict our analysis to $\hat{\nu}^2 \geq 1$. Since the metric is continuous in $\hat{\nu}$, we can use the previous results obtained for the asymptotically G\"odel geometries \cite{Compere:2007in}. All kinematical aspects of this analysis will be valid in the case $\hat{\nu}^2 \geq 1$ by continuity. We simply have to adapt the dynamical aspects of the discussion to TMG by considering a gravitational Chern-Simons interaction in place of a coupling to the Maxwell-Chern Simons theory. 

We consider in this paper only the phase space of black hole solutions \eqref{BH}. More general phase spaces can be constructed but our analysis of central charges applies already in that simplified setting. Following \cite{Compere:2007in}, the spacelike streched black hole geometries can be shown to admit the following two sets of asymptotic Killing vectors ($n \in \mathbb Z$)
\begin{eqnarray}
l_n &\equiv & (N e^{i n \Phi}+O(1/R))\d_T +(-i \,n\, R e^{i n \Phi}+O(R^0))\d_R \nonumber \\
&& +(e^{i n \Phi}+O(1/R^2))\d_\Phi,\label{norm}\\
t_n &\equiv & (N^\prime e^{i n \Phi}+O(1/R))\d_T .
\end{eqnarray}
All associated charges can be shown to be finite and conserved in the phase space. The normalizations $N$ and $N^\prime$ are solution-dependent factors that are left unfixed by the asymptotic analysis. They are however constrained by the fact that the conserved charges should be integrable in the phase space. These asymptotic symmetries form the algebra
\begin{eqnarray}
i[l_m,l_n] &=& (m-n)l_{m+n}, \nonumber\\
 i[l_m,t_n]&=& -n t_{m+n},\qquad [t_m,t_n] = 0.\label{alg}
\end{eqnarray}
The subset $(l_{-1},l_0,l_1,t_0)$ form a $sl(2,\mathbb R) \oplus \mathbb R$ subalgebra. Let us denote the charges differences between the black hole metric $g$ and the background $\bar g$ ($r_+=r_-=0$) by $L_n \equiv \cQ_{l_n}[g;\bar g]$, $T_n \equiv \cQ_{t_n}[g;\bar g]$. 

One can check using the definition of $E^{\mu\nu}$ \eqref{defE} in that the technical assumption (4.3) of \cite{Barnich:2007bf}, or more simply, the sufficient condition (2.7) of \cite{Compere:2007in} is satisfied in the phase space \eqref{BH}. The term $E^{\mu\nu}$ simply does not contribute to any charge. Therefore, the charges form a representation of the asymptotic symmetry algebra. One can then plug in the algebra \eqref{alg} into the first term on the right-hand side of \eqref{eq:repr} and use the linearity of the charges with respect to the Killing vector to factorize the $m,n$ factors. The central charge $\cK_{\xi,\xi^\prime}[\bar g] \equiv \int_S k^{\mu\nu}_{\xi}[\cL_{\xi^\prime} \bar g;\bar g]$ can be computed by implementing the formula \eqref{chargetot} in a Mathematica code. We find the following centrally extended Virasoro algebra
\begin{eqnarray}
i \{L_m,L_n\} &=& (m-n)L_{m+n}+ \frac{c}{12}m(m^2- \bar\beta )\delta_{m+n,0}, \nonumber \\
 i\{L_m,T_n\} &= & -n T_{m+n} - \frac{3+\hat \nu^2}{12\hat \nu l G}\bar N \bar N^\prime\,m\delta_{m+n,0},\nonumber\\
i\{T_m,T_n\} &=& -\frac{3+\hat \nu^2}{12\hat \nu  l G} \bar N^\prime \bar N^\prime\,m\delta_{m+n,0}
\end{eqnarray}
where $\bar N$ and $\bar N^\prime$ are normalization factors of the generators, see \eqref{norm}, on the background, $\bar \beta = \frac{(\hat{\nu}^2+3)^2 \bar N \bar N}{l^2(5 \hat{\nu}^2+3)}$, and 
\begin{eqnarray}
c = \frac{(5 {\hat{\nu}}^2 + 3)l}{G {\hat{\nu}}({\hat{\nu}}^2 + 3)}
\end{eqnarray}
is the Virasoro central charge. This central charge is the value $c_R$ conjectured in \cite{Anninos:2008fx} \footnote{In the versions v1 and v2 of this paper, the central charge had an incorrect minus sign due to a minus sign mistake in the expression for $\mathcal K_{\xi,\xi^\prime}$ given after equation (6).}. We leave open the question of whether a Virasoro algebra with central charge $c_L= \frac{4\hat \nu l}{G(\hat \nu^2+3)}$ and commuting with $\{L_n\}_{n\in \mathbb Z}$ can be constructed via a Sugawara-type procedure from the current algebra $\{ T_n \}_{n\in \mathbb Z}$.

We will now show that $L_0$ is bounded from below. The variation of charge between two black hole solutions of parameters $(\nu,J)$ and $(\nu+\delta \nu,J+\delta J)$ is obtained from $\delta L_0 \equiv \int_S k^{\mu\nu}_{l_0}[\delta g;g]\eps_{\mu\nu\rho}dx^\rho$ as 
\begin{eqnarray}
\delta L_0 = \frac{2}{3\alpha}(2 G\nu +\alpha N)\delta \nu - \frac{1+3\alpha^2 L^2}{3\alpha^2 L^2}\delta J.
\end{eqnarray}
It turns out that there is a natural ansatz for $N$. The black hole solutions admit horizons only under the condition $2 \nu^2 - \frac{1+\alpha^2 L^2}{\alpha L^2} \frac{J}{G} \geq 0$ \cite{Banados:2005da}. Choosing $N = \frac{4\alpha L^2}{(1+\alpha^2 L^2)}G \nu$ and integrating in the phase space between $(\nu=J=0)$ and $(\nu,J)$, we obtain
\begin{eqnarray}
L_0 = \frac{1+3\alpha^2 L^2}{3\alpha (1+\alpha^2 L^2)}\left( 2G \nu^2 - \frac{1+\alpha^2 L^2}{\alpha L^2} J \right) \geq 0
\end{eqnarray}
for regular black holes and $G>0$. Remark that exactly the same anzatz was used in \cite{Compere:2007in}, and a non-negative spectrum for $L_0$ though different from an overall positive factor was obtained.

We have therefore shown that when the black holes have a non-negative $L_0$ spectrum, the Virasoro central charge is also positive \footnote{In v1 and v2 of this paper, this fact was incorrectly disclaimed. The change is this conclusion is a consequence of the sign mistake explained in [41].}. One may want to invert the sign of the Newton constant as done e.g. in \cite{Carlip:2008jk,Carlip:2008eq}. With $G < 0$, the central extension of the $L_n$ generators becomes negative and the regular black holes have a mass spectrum $L_0$ unbounded from below. In that case, one can however consider the generators $\tilde L_{n}=-L_{-n}$ for which both the central charge and the zero mode $\tilde L_0$ are non-negative. 

Our result is in favor of the conjecture that spacelike warped geometries have a regular CFT dual \cite{Anninos:2008fx} \footnote{This conclusion has been changed in v3, see footnotes  [41,42].}. It would be interesting to compute the linearized energy spectrum of perturbations around the spacelike warped geometries to confirm the stability of the background.

\section*{Acknowledgments}

GC thanks the string theory group of Milano for their hospitality and support. We thank G. Clement for his remarks on his paper and G. Giribet for his comments and encouragements. Useful discussions and exchanges with S. Bestiale, M. Guica, T. Hartman, M. Leoni, W. Li, R. Olea, G. Tagliabue, W. Song and A. Strominger on some topics dealt with in this paper are greatly acknowledged. GC sincerely thanks T. Nishioka and K. Murata for pointing out the reason of the sign mistake in the central charge in v1-v2. This work was supported in part by the US National Science Foundation under Grant No.~PHY05-55669, by funds from the University of California, by INFN and by the Italian MIUR-PRIN contract 20075ATT78. GC was supported as David and Alice van Buuren Fellow of the BAEF foundation.


\end{document}